\documentclass[preprint,floats,aps,epsfig,nofootinbib,amssymb]{revtex4-1}
\usepackage{mathrsfs}
\usepackage{slashed}
\usepackage{graphicx,color}
\usepackage{subfigure}
\usepackage{dcolumn}
\usepackage{bm}
\usepackage{subfig}
\usepackage{graphicx}
\usepackage{mathrsfs,amsmath,amssymb}
\usepackage{stackrel}
\usepackage{bbding}
\usepackage{pgfplots}
\usetikzlibrary{positioning}
\usetikzlibrary{snakes}
\def \bea {\begin{eqnarray}}
\def \eea {\end{eqnarray}}

\begin{document}

\title{Probing Sub-eV  Dark Photon, Scalar and Axion-like Particle Dark Matters with Transmon Qubits}

\author{Wei Chao$^{1,2}$}
\email{chaowei@bnu.edu.cn}

\author{Yu Gao$^3$}
\email{gaoyu@ihep.ac.cn}

\author{Ming-Jie Jin$^{4}$}
\email{202411006@jcut.edu.cn}

\author{Xiao-Sheng Liu$^{1,2}$}

\author{Xi-Lei Sun$^5$ }
\email{sunxl@ihep.ac.cn}
\affiliation{
$^1$Center of Advanced Quantum Studies, School of Physics and Astronomy, Beijing Normal University, Beijing, 100875, China \\
$^2$Key Laboratory of Multi-scale Spin Physics, Ministry of Education,
Beijing Normal University, Beijing 100875, China \\
$^3$Key Laboratory of Particle Astrophysics, Institute of High Energy Physics, Chinese Academy of Sciences, Beijing, 100049, China\\
$^4$School of Mathematics and Physics Science,
Jingchu University of Technology, Jingmen, 448000, China\\
$^5$State Key Laboratory of Particle Detection and Electronics, Institute of High Energy Physics, Beĳing, 100049, China
}
\vspace{3cm}

\begin{abstract}

In this paper, we investigate constraints of the transmon qubit, an improved version of the charge qubit, on bosonic light dark matters. Phonon excitations induced by the scattering or absorption of dark matter on a superconductor may destroy the Cooper pair, leading to the production of quasiparticles made by the electron. By measuring the production rate of the quasiparticle density, one may read out the coupling between dark matter and ordinary matter, assuming that these quasiparticles are solely induced by dark matter interactions. For the first time, we show constraints on the parameter space of the dark photon, light scalar dark matter, and axion-like particles from the measurement of quasiparticles in transmon qubit experiments. This study offers insights for the development of quantum qubit experiments aimed at the direct detection of dark matter in underground laboratories.

\end{abstract}

\maketitle
\section{Introduction}
Various astrophysical and cosmological observations have confirmed the existence of cold dark matter (DM)~\cite{Bertone:2004pz,Planck:2018vyg}, which should be a non-baryonic, massive, and weakly interacting particle; there is no cold DM candidate in the standard model (SM) of particle physics, triggering the development of DM models. Among various promising DM candidates~\cite{Cirelli:2024ssz}, the foremost task is to detect DM signals in the laboratory.

In recent years, direct detection technologies for DM~\cite{Billard:2021uyg,Schumann:2019eaa,MarrodanUndagoitia:2015veg,Fitzpatrick:2012ix,Essig:2011nj} have developed rapidly. The exclusion limits provided by experiments utilizing nuclear recoil technology have been continuously strengthened, and numerous experiments employing electronic excitation have been conducted to probe DM, enabling the detection of lighter DM candidates. However, these experiments are mostly limited to non-relativistic DM candidates with a mass above the MeV scale. Furthermore, when considering the DM absorption process, the detection threshold can be lowered to the electroVolt (eV) level. In fact, there are also sub-eV DM models with strong theoretical motivations, such as the axion and the dark photon. These particles can be produced in the early universe through mechanisms like the misalignment mechanism and parametric resonance, etc.
Direct Detection of ultralight DM~\cite{Hui:2016ltb,Dror:2024con,Fukuda:2018omk,Sun:2024qis} is an important unresolved issue, and various proposals have been put forward to explore sub-eV DM. Electronic excitations in targets with small excitation gaps can be applied to the detection of light DM~\cite{Essig:2017kqs}. The expected target materials include Dirac materials~\cite{Hochberg:2017wce}, superconductors~\cite{Hochberg:2015pha}, and narrow-gap semiconductors~\cite{Hochberg:2016sqx}.
%
%
The emergence of collective excitations like phonon~\cite{Kahn:2021ttr,EDELWEISS:2023hcg,Campbell-Deem:2022fqm,Lyon:2022sza,Coskuner:2021qxo,Cox:2019cod,Chao:2023liu}, magnon~\cite{Trickle:2019ovy,Mitridate:2020kly,Chigusa:2020gfs,Chigusa:2023bga,Barbieri:1985cp,Flower:2018qgb} and plasmon~\cite{Liang:2024xcx,Liu:2023rmu,Kozaczuk:2020uzb}, resulting from DM-matter scattering or DM absorption within condensed matter material, represents a promising avenue for DM direct detection. This is particularly relevant given that these excitations possess energies within the ${\cal O}(1\sim100)$  meV range, aligning well with the energy scales characteristic of light DM.
%
%
In brief, researchers are actively exploring and employing a variety of methods to detect DM, yet no definitive detection has been achieved. Utilizing new materials or technologies to detect light DM presents an increasingly recognized challenge in the fields of high-energy physics and precision measurement physics. The path to the direct detection of DM remains a lengthy and challenging journey.
%

In this study, we delve into the potential constraints imposed on light DM by superconducting qubits~\cite{Dixit:2020ymh}. These quantum qubits, constructed from superconducting materials and maintained at ultra-low temperatures, have garnered significant interest among quantum computing scientists for their enhanced coherence and stability.
Recently, transmon qubits~\cite{Rist__2013} as DM detectors have been proposed.  Refs.~\cite{Chen:2022quj,Chen:2024aya} have showed how to detect hidden photon and the QCD axion using the direct excitation of transmon qubits. Refs.~\cite{Das:2022srn,Das:2024jdz} have shown constraint on the DM-nucleon scattering cross sections using the latest transmon qubit measurements~\cite{Dixit:2020ymh}.  In addition, a transmon qubit prototype for DM detection experiment has been designed in Ref.~\cite{Moretti:2024xel}.  
Compared with conventional direct detection experiments, the transmon qubit facility have several merits on the direct detection of light DM: (1) Transmon qubits maintain long quantum coherence up to microseconds at cryogenic temperatures, enabling precise detection of faint energy signals from DM interactions; (2) Operated in dilution refrigerators, transmon qubits benefit from minimal thermal and electromagnetic noise, suppressing interference that could obscure DM signals, which is critical for detecting DM interactions with extremely low energy threshold; (3) The energy level of transmon qubits  can be precisely adjusted using external microwaves or magnetic fields, adapting to different DM models with specific couplings. In addition, arrays of transmon qubits on a single chip can simultaneously detect DM signals across multiple frequency bands, enhancing detection efficiency and statistical significance.  All these advantages make transmon qubit experiments a potential DM direct detection approach worthy of continuous investigation, especially for probing light DM.  
In this paper, We focus on the case of the absorbable DM. Light bosonic DM can be absorbed by the superconducting material into phonons via  DM-proton or DM-neutron interactions and the absorption rate can be calculated utilizing the effective field theory developed in Ref.~\cite{Mitridate:2023izi}.  These phonon can break Cooper pairs whenever their energy is above meV, leading to to the production of quasiparticles.  With the help of transmon qubit measurements~\cite{Dixit:2020ymh}, we put constraints on the DM couplings with results shown in Figs.~\ref{figure1} and \ref{figure2}, where constrains on the dark photon, light scalar DM and axion-like particle DM are presented. These results are the first solid constraints on light bosonic DM from quantum qubits experiments, and will insights for the development of quantum qubit experiments aimed at the direct detection of dark matter in underground laboratories. 
 
The remaining of this paper is organized as follows: In section II we briefly review the local DM density near the sun and present specific bosonic DM models. Section III is devoted to the numerical study of constraint on various DM couplings. The last part is concluding remarks.  

\section{Light Bosonic DMs}

Although the particle nature of DM remains elusive, astronomical observations offer good estimates of the relic density of the Universe and the nearby DM density in the vicinity of the Solar system. 
DM density near the Sun can estimated from stellar kinematics within the Galaxy. The commonly used local DM density near the Solar system is approximately $0.4~{\rm GeV/cm^3}$~\cite{Pato:2015dua}, which has an ${\cal O}(1)$ uncertainty and it may vary in the range $\rho_{\rm DM} =(0.2-0.6) ~{\rm GeV/cm^3}$~\cite{Read:2014qva}. Alternatively, in case DM interacts strongly with the SM particles, DM may become captured by celestial objects, including the Earth, resulting in a much larger local density than the average density of the neighborhood ~\cite{Neufeld:2018slx}. 

 
In this paper, we study phonon signals that arise from the absorption of light bosonic DM particles, specifically the dark photons (DP), light scalars, and axion-like particles (ALPs). 
 %
%
First we can consider the dark photon with its low-energy Lagrangian characterized by a kinetic mixing with the SM~\cite{Holdom:1985ag,Fabbrichesi:2020wbt,An:2014twa}:
\bea
{\cal L}_{\rm DP}^{} \supset -\frac{1}{4} F_{\mu\nu}^{} F^{\mu\nu}_{} -\frac{1}{4}F^\prime_{\mu\nu} F^{\prime\mu\nu}_{}  +\frac{\kappa}{2}F_{\mu\nu}^{} F^{\prime\mu\nu}_{}+ \frac{1}{2} m_X^2 X_\mu^{} X^\mu_{} 
\eea
where $X_\mu$ is the DP field, $F_{\mu\nu}^{}$ and $F^\prime_{\mu\nu}(=\partial_\mu^{} X_\nu^{} -\partial_\nu^{} X_\mu^{})$ are the field strength tensors for the SM photon and the DP, $m_X$ is the DP mass, and $\kappa$ is the kinetic mixing parameter. There are several interesting DP production mechanisms that may address the correct abundance of DM.  One of the simplest ways to produce DP is the misalignment mechanism~\cite{Nelson:2011sf},  another one is based on the tachyonic instability arising when the DP couples to a misalignment axion~\cite{Co:2018lka}, and a third one is from the decay of cosmic strings~\cite{Long:2019lwl}.  DP on the Earth, may come from the burning of the Sun with the DP flux proportional to the photon flux rescaled by a factor $\kappa^2$, or from the relic density of DM, which is about $0.4~{\rm GeV/cm^3}$. The absorption of DP by a superconductor deposits energies in the term of phonon,  which may subsequently break the Cooper pair. This effect will be investigated in the next section.

Secondly, we consider a bosonic DM model of a light scalar particle, $\phi$, with its interaction to the electrons and nucleon given by
\bea
{\cal  L } \supset  \sum_{f =e, p, n} d_{\phi ff}  \phi \bar f f 
\label{eq:boson_DM}
\eea
Similar to the DP model, the relic abundance of the scalar DM can also be provided by the misalignment mechanism.  There is also  modified definitions of the coupling between the scalar DM and SM fermions. Note there are alternative definitions of the coupling in literature. For instance, Refs.~\cite{Adelberger:2003zx,Arvanitaki:2015iga} use a normalization that rescales the coupling $d_{\phi ff} \to d_{\phi ff} \sqrt{4\pi} m_f /M_{\rm pl}$ in the Eq.~\ref{eq:boson_DM}, where $m_f$ and $M_{\rm pl}$ are the fermion mass and Planck mass scale, respectively. 

The third model to be studied is the ALP,  a generic prediction of many high energy physics models including string theory.  The ALP is one of the most theoretically motivated DM candidate since it may also address the strong CP problem. The ALP  DM can be generated in the early universe via the misalignment mechanism. The predicted mass range for the QCD axion is $10^{-6}~{\rm eV}\leq m_a \leq 10^{-5}~{\rm eV}$, while the preferred mass range of the ALP can be  larger since it do not necessarily solve the strong CP problem.  In this paper we will focus on the the derivative ALP couplings
\bea
{\cal L} \supset \sum_{f=e, p,n} \frac{g_{aff}^{} }{2 m_f^{} } \partial_\mu^{} a \bar f \gamma^\mu \gamma^5 f 
\eea
where $m_f$ is the fermion mass. Given these interactions, an ALP can be absorbed  by the electron, proton or neutron when it pass through a target, producing phonon signal.  Free parameters in this process includes: $m_a$, $g_{aff}$ and the ALP flux.  The ALP flux may come from the cosmic relic abundance controlled  by  $\rho_{\rm DM}$ and its velocity distribution. Alternatively it can be produced in stars~\cite{Carenza:2024ehj}, especially in the sun and other nearby stars.   In this study we will not consider DM flux emitted from the sun, whose impact to the quasi-particle will be studied in a future work. On the other hand we take the dark matter flux as $\rho_{\rm DM}^{} v$ with $v$ the velocity of DM and $\rho_{\rm DM} (\sim 0.4 ~GeV/cm^3 )$  the local DM energy density.

\section{Detectability at transmon qubits}
\label{sect:detectability}
In recent years, the global pursuit of quantum advantage~\cite{Arute:2019zxq,Zhong:2020iql,Wu:2021kmx} has escalated the competition in the development of quantum calculation devices. This race encompasses a diverse range of technologies, with circuit quantum electrodynamics (cQED) devices [56] at the forefront of research and innovation. Among these  devices, the transmon qubit has emerged as one of the most pivotal components across various quantum computing architectures, owing to its remarkable performance and scalability. A transmon qubit~\cite{Rist__2013} represents an advanced iteration of the charge qubit, also known as a Cooper-pair box. Its design features a Josephson junction—a fundamental element consisting of two superconducting layers separated by a thin insulating barrier—connected in parallel with a large shunting capacitor~\cite{Makhlin:2001zz}. This configuration addresses the inherent limitations of traditional charge qubits, particularly their susceptibility to charge noise. By introducing the shunting capacitor, the transmon qubit effectively suppresses fluctuations in electrical charge, thereby stabilizing its quantum state. One of the transmon qubit's improtant attributes is its extended coherence time, a parameter that quantifies how long the qubit can maintain its quantum state without decohering or losing information. This characteristic makes transmon qubits indispensable for building practical quantum computers capable of tackling complex problems beyond the reach of classical machines. Beyond quantum computing, the transmon qubit's resilience to charge noise makes it as an ideal candidate for DM direct detection experiment~\cite{Chen:2022quj,Chen:2024aya,Das:2022srn,Das:2024jdz,Moretti:2024xel}. Transmon qubits can serve as highly sensitive detectors, leveraging their ability to detect minute energy exchanges or quantum state perturbations that may result from interactions with DM particles. Their low-noise operation and precise quantum state control enable the identification of faint signals that would otherwise be obscured by background noise, thus opening up new avenues for exploring the nature of the universe's elusive DM component.

In this section we follow the strategy of Refs.~\cite{Das:2022srn,Das:2024jdz} to investigate the indirect detection signal of DM by measuring the quasi-particle density induced by phonons which come from  the scattering or absorption a low mass DM particle on the facility. Ref.~\cite{Rist__2013} has investigated the decoherence of a transmon qubit, using a single-junction superconducting qubit made by aluminum, and the quasi-particle density is measured as $n_{qp} \approx 0.04\pm 0.01~ \mu m^{-3}$. It is well-known that low temperature allows electrons to form Cooper pairs, bound via long range interactions with phonons. When DM scatters with the superconducting aluminum, its energy is partially deposited in the form of phonons, which can break Cooper pairs if phonons have energy exceeding the binding energy of Cooper pairs and lead to an excess of quasiparticles. 
The steady-state quasiparticle density $n_{qp} $ can be estimated using the mean field results~\cite{PhysRevLett.117.117002}, 
\bea
\frac{d n_{qp}}{dt} = -\Gamma_R -\Gamma_T + \Gamma_G
\eea
where $n_{qp}$ is the quasi-particle density, $\Gamma_R$, $\Gamma_T$ and $\Gamma_G$ are the recombination rate, trapping rate and generation rate, respectively.  The recombination and trapping rates can be written as $\Gamma_R \sim \bar \Gamma n_{QP}^2$ and $\Gamma_T = \bar \Gamma_T n_{\rm QP}$, while $\Gamma_G$ is related to the interaction of DM with the superconductor material, which will be discussed in the following.  In our analysis,  we take $\bar \Gamma =40~\mu m^3 /s$ for aluminium~\cite{PhysRevLett.117.117002} and take $\bar \Gamma_T=0$~\cite{Das:2022srn,Das:2024jdz},  for simplicity. 

For quasiparticles in equilibrium, one has $d n_{pq}/dt =0$, and thus
\bea
\Gamma_G = \bar \Gamma n_{\rm QP}^2+ \Gamma_T n_{\rm QP} \label{master1}
\eea
which can be used to determine the DM-superconductor scattering rate or absorption rate given the value of $n_{\rm QP}$. On the other hand, the quasiparticle generation rate can be written as~\cite{Das:2022srn,Das:2024jdz}\footnote{ Here we have included an extra  factor $ \rho_T$ in the formula, considering that $R$ is DM interaction rate per unit time per unit target mass. }
\bea
\Gamma_G= \frac{\zeta}{2\Delta} \int d \omega \omega \frac{d R }{d \omega} \rho_T \label{master2}
\eea
where $\zeta=0.6$~\cite{Hochberg:2021ymx,PhysRevB.14.4854} being the quasi-particle generation efficiency, and $\Delta=340~{\mu eV}$ denoting the superconducting energy gap of aluminum, $R$ is DM scattering or absorption rate per unit time per unit target mass, $\omega$ is the frequency of the phonon and $\rho_T$ is the density of the target.  Combing Eqs. (\ref{master2}) and (\ref{master1}), one may derive a constraint on the DM coupling strength for a given DM scattering rate or absorption rate. In the following, we will consider constraints of transmon qubit measurements on low-mass DM candidates including an ALP, a scalar DM and a dark photon, by deriving limits on their couplings. 

\begin{figure}[t]
\includegraphics[width=0.45\textwidth]{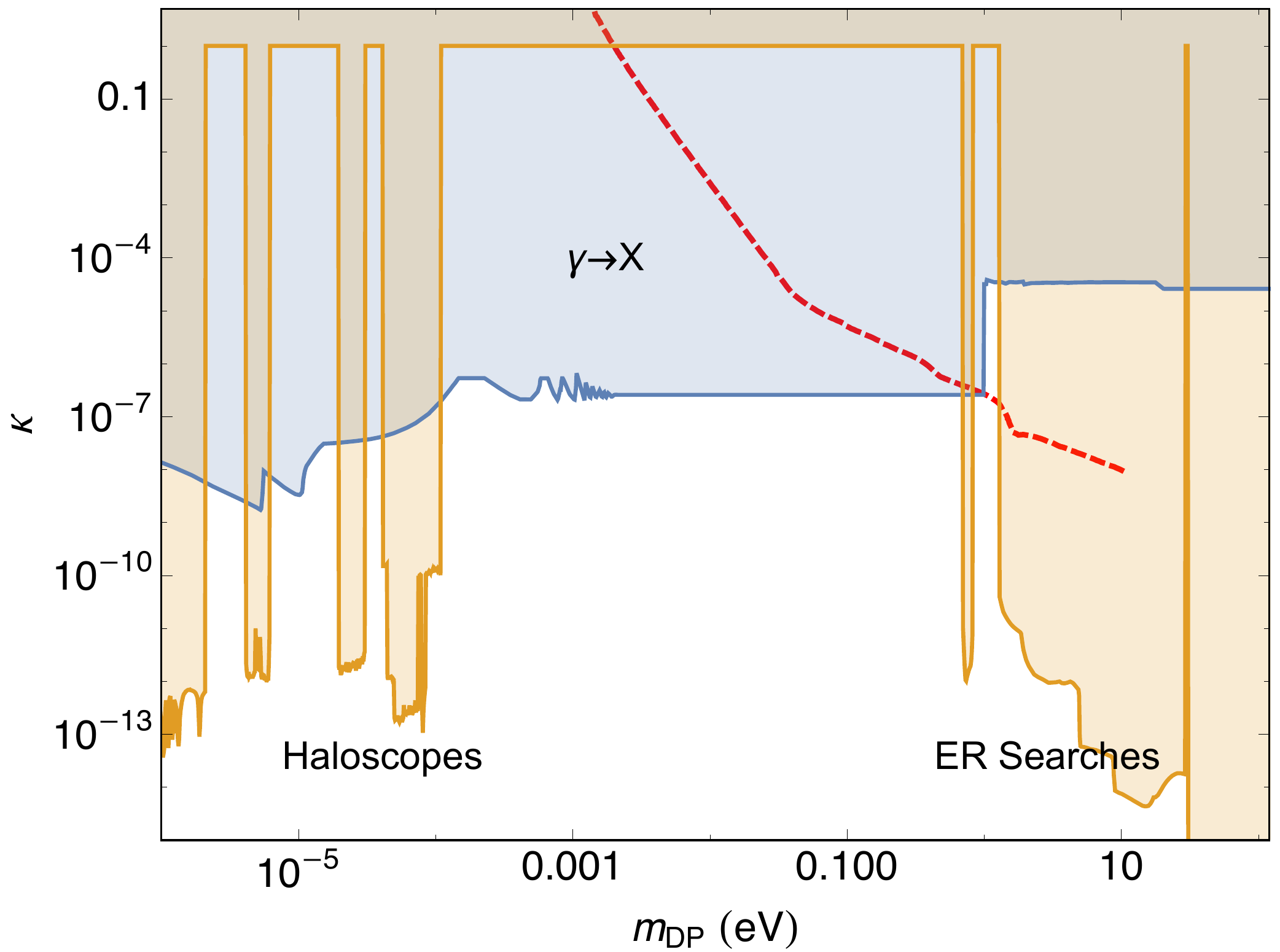}
\hspace{0.5cm}
\includegraphics[width=0.46\textwidth]{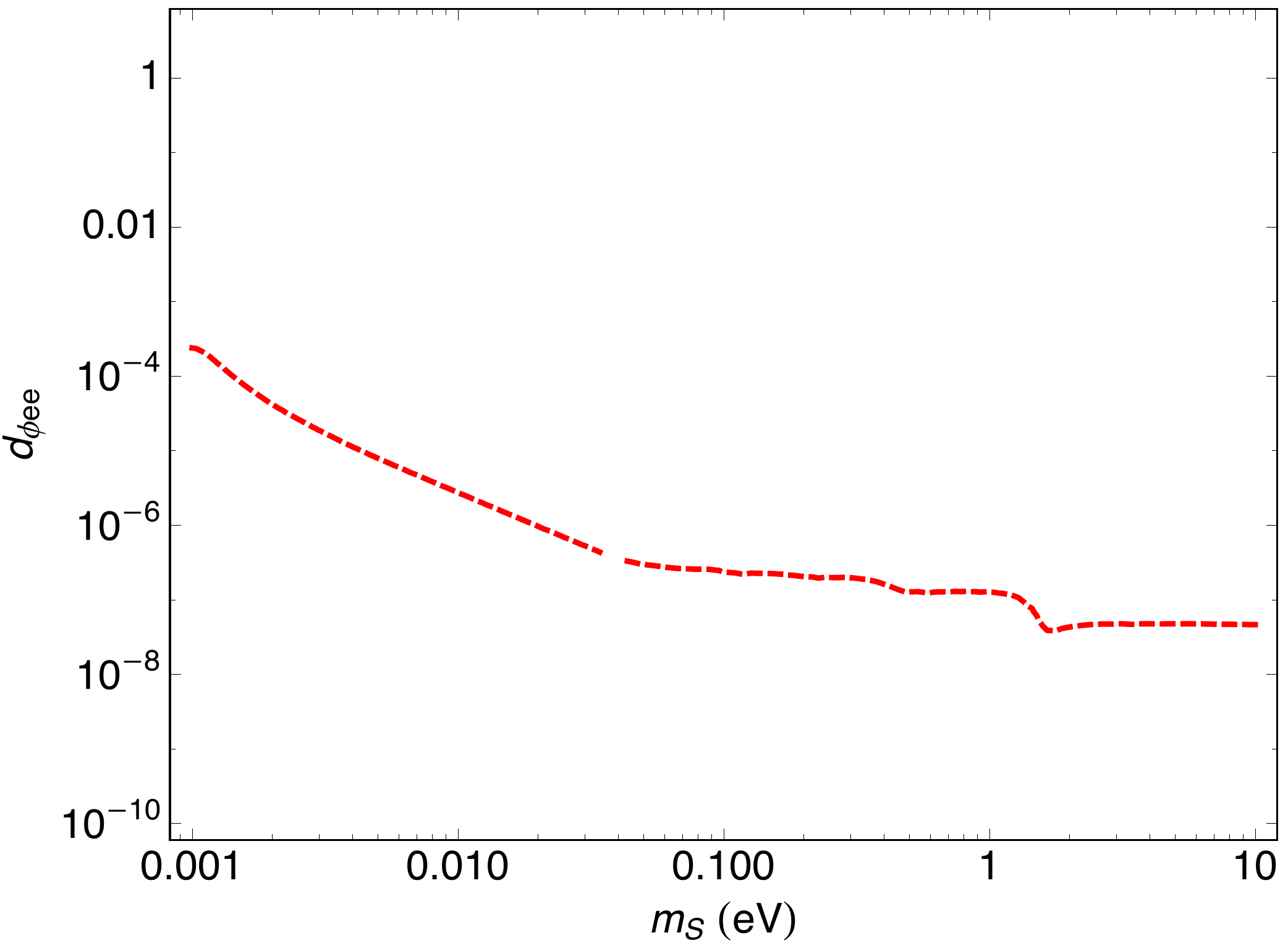}
\caption{Left-panel: Projected sensitivity on the dark photon mixing $\kappa$ from the transmon qubit, based on a sensitivity $n_{\rm qp}\le 0.04 ~{\rm \mu m}^{-3}$. Existing DP-photon conversion limits (blue-shaded) and direct laboratory search results from haloscope and electron recoil experiments (light-brown)~\cite{Caputo:2021eaa} are shown for comparison. Right-panel: projected sensitivity on the scalar dark matter coupling $d_{\phi ee }$.}
\label{figure1}
\end{figure}

In order to put constraints on absorbable bosonic DM using the  measured quasi-particle density in a transmon qubit facility, one needs to first calculate the energy deposition induced by the absorption of light bosonic DM. Recent studies ~\cite{Hardy:2016kme,Mitridate:2021ctr,Chen:2022pyd,Krnjaic:2023nxe,Mitridate:2023izi} have shown that the DM absorption rate can be written in terms of in-medium self-energies by using the optical theorem.  The absorption rate of the $\lambda$th polarization of the DM, $\chi$, can be written as
\bea
\Gamma^\lambda = -\frac{1}{m_\chi} {\rm Im} [\Pi_{\chi \chi}^\lambda]
\eea 
where $m_\chi$ is the DM mass, $\Pi_{\chi\chi}^\lambda$ is the self-energy between two $\chi$ particles of the $\lambda$th polarization. $\Pi_{\chi\chi}^\lambda$ receive contributions from both electronic excitations and phonon excitations~\cite{Mitridate:2023izi}.  Assuming the electronic band gap is much larger than the energy of phonon excitations and taking use of real part of the electron self-energy given in Refs~\cite{Mitridate:2021ctr,Chen:2022pyd,Krnjaic:2023nxe}, one only needs to consider phonon contribution to self energies, which has been calculated in the Ref.~\cite{Mitridate:2023izi}.

\subsection{Dark photon}

As a result, the total absorption rate  of the dark photon dark matter per unit target mass  per unit time can be written as  
\bea
R_{\rm DP}^{}  = \frac{\rho_{\rm DP}^{}}{\rho_T} \kappa^2  {\rm Im } \left[ - \frac
{1}{\varepsilon(\omega)}\right] \label{dpmaster}
\eea
where $\rho_{\rm DP}^{}$ is the local dark photon energy density. $\rho_T$ is the target energy density that depends on the material. For aluminum, $\rho_T$ is typically $2.7 g/cm^3$.  $\kappa$ is the DP-photon kinetic mixing parameter given in the Eq. (1), and the last term is the energy loss function (ELF), which is evaluated using the public code {\bf  DarkELF}~\cite{Knapen:2021bwg}, $\varepsilon(\omega)$ is the frequency dependent longitudinal dielectric function.

Taking  Eq.~(\ref{dpmaster}) into Eqs. (5) and (6),  one is able to derive exclusion limits on the kinetic mixing parameter, $\kappa$.  We show in the left-panel of the Fig.~\ref{figure1}, the  contour in the $m_{\rm DP}-\kappa$ plane that may address the observed density of quasi-particles, where $m_{\rm DP}$ is the DP mass.  It should be mentioned that these quasi-particles may also come from the thermal population, $n_{qp}^{tot} =  n_{qp}^{\rm DM} + n_{qp}^{th}$, with~\cite{Das:2024jdz}
\bea
n_{qp}^{th} = 2 \nu_0 \Delta \sqrt{\frac{2\pi k_B T}{\Delta} } \exp\left( -\frac{\Delta }{k_B T} \right)
\eea  
where $\nu_0=1.2\times 10^4 ~\mu m^{-3} \mu eV^{-1}$ being the Cooper pair density at the Fermi level in the aluminum, $k_B$ is the Boltzmann constant.  It means that the curve in the plot only gives the exclusion limit on the parameter space: regime above the curve is excluded. It should be mentioned that the thermal quasiparticle density at a temperature $T$ is indeed a key parameter setting the baseline for any detection experiment. Our assumptions for the thermal background are conservative and align with these established experimental results. We fully agree that any future dedicated dark matter detector would require in-situ characterization of its specific quasiparticle background, the constraints we derive are based on experimentally validated physics. The projected sensitivity for a future optimized device would, of course, be contingent upon achieving thermal quasiparticle densities at or below the levels already demonstrated in state-of-the-art qubit experiments.

 
\begin{figure}[t]
\includegraphics[width=0.45\textwidth]{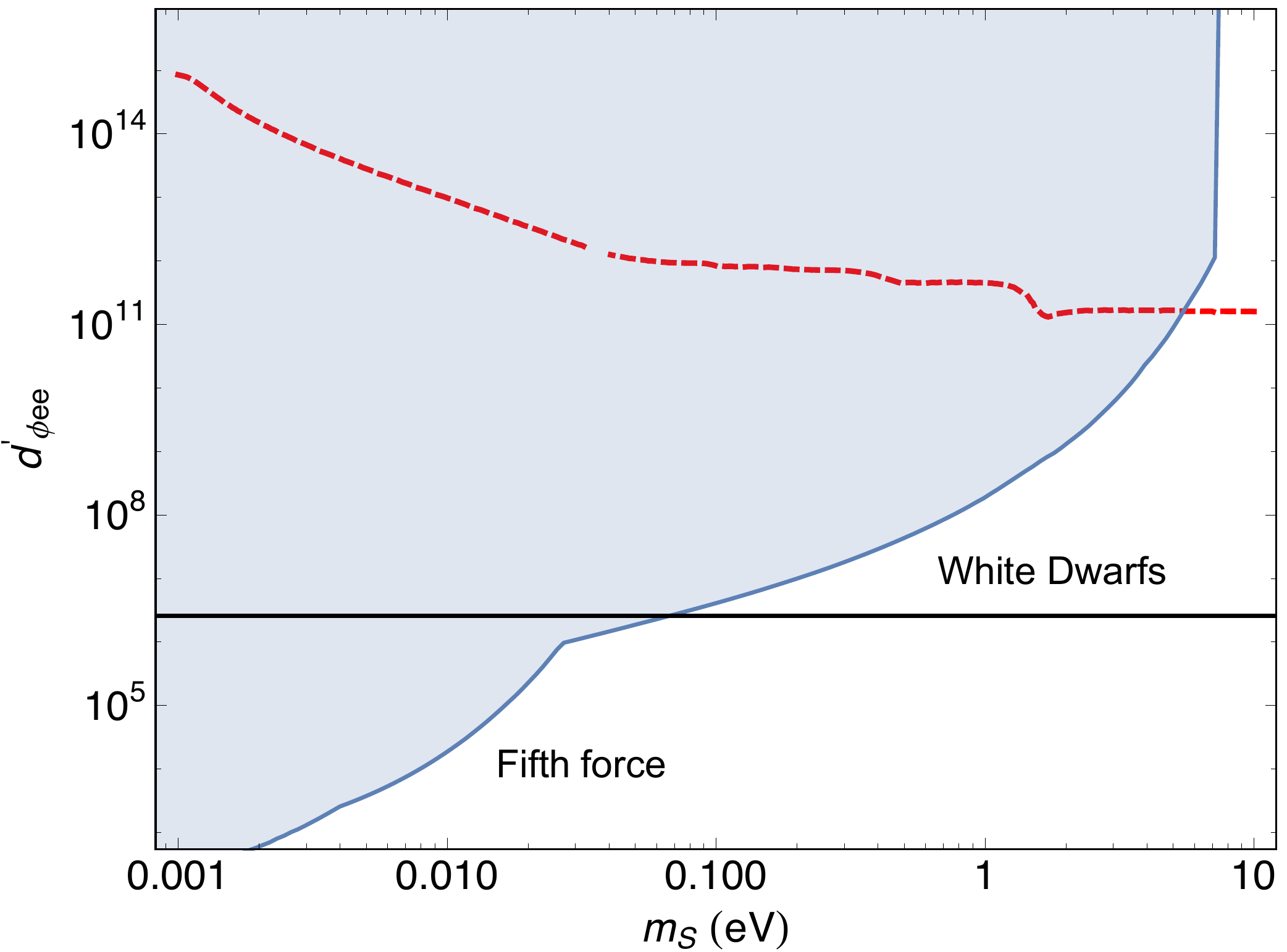}
\hspace{0.5cm}
\includegraphics[width=0.45\textwidth]{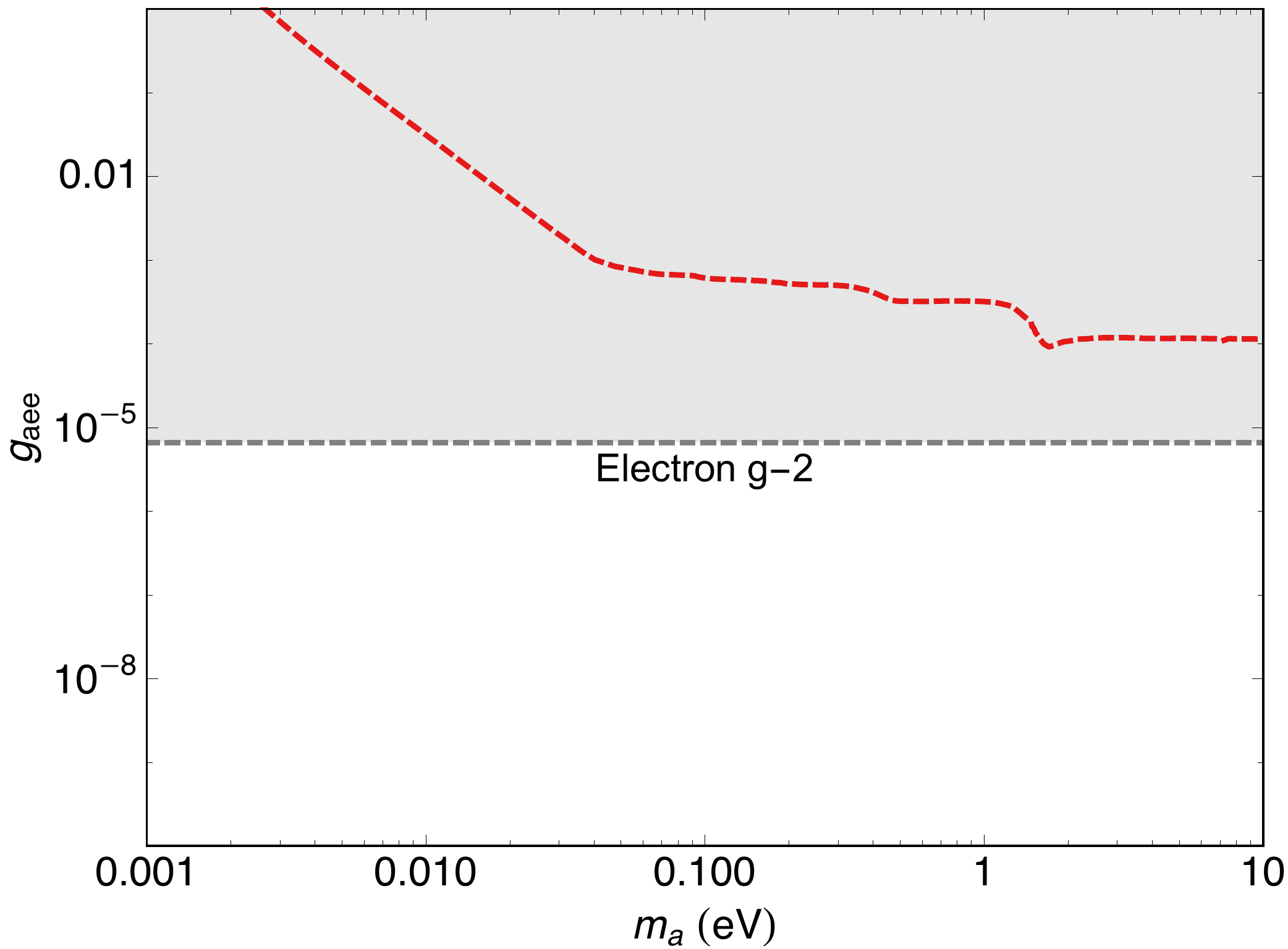}
\caption{Left-panel: Transmon qubit constraint on the light scalar DM, where the light-blue regime is excluded by the fifth force measurement~\cite{Adelberger:2003zx} and horizontal line is the limit of the White Dwarfs observations~\cite{Bottaro:2023gep}. Right-panel: Transmon qubit constraint on the axion-like particle in the $m_a-g_{aee}$ plane (solid), based on a sensitivity estimate with $n_{\rm qp}\le 0.04 ~{\rm \mu m}^{-3}$. The limit from measurements on the electron dipole moment $g_e-2$ is shown for comparison (red shaded). Other constraints are not included considering that $g_e-2$ results encompass all constraints of transmon qubit.}\label{figure2}
\end{figure}

\subsection{Scalar DM}
For the scalar DM $\phi$, the absorption rate can also be derived from the optical theorem by evaluating the self-energies in the medium. Here we adopt the assumption in  Ref.~\cite{Mitridate:2023izi} that the coupling $d_{\phi ff}$ is photon-like, and the total absorption rate can be given in the term of the ELF~\cite{Mitridate:2023izi},
\bea
R\sim {d_{\phi ee}^2 \over e^2} {q^2\over \omega^2 } {\rho_{\phi}^{} \over \rho_T} {\rm Im} \left[ -\frac{1}{\varepsilon (\omega)}\right]
\eea 
where $q$ and $\omega $ are the momentum and energy of the incoming scalar DM, respectively. For cold dark matter, $q/\omega \sim 10^{-3}$ after virialization. Alternatively, for relativistic $\phi$ produced from the center of the Sun, $q/\omega \sim 1$.  Here we will consider $\phi$ as the major component of cold dark matter and take $\rho_\phi=0.4~ {\rm GeV/cm^3}$.  

We show in the right-panel of the Fig.~\ref{figure1}   constraints on the scalar DM in the  $d_{\phi ee}-m_\phi$ plane. Again, the region above the contour is excluded.  To compare with other constraints from stellar cooling~\cite{Bottaro:2023gep} and fifth force~\cite{Adelberger:2003zx}, one needs to re-parameterize the coupling, $d^\prime_{\phi ee} =d_{\phi ee} M_{pl}/(\sqrt{4\pi} m_e)$, where $M_{pl}$ is the Plank mass.  Constraints on the new coupling is listed in the left-panel of the Fig.~\ref{figure2}.

\subsection{Axion-like particle}
ALP is one of the most motivated wave-like dark matter candidates. To evaluate the detectability of the transmon qubit on the ALP, we consider its derivative coupling to the electron's axial current $g_{aee}$. To simplify the calculation of the absorption rate, we assume a naive ALP model in which $g_{aee}/m_e=-g_{app}/m_p $ and  $g_{ann}=0$. With this parameter choice the ALP absorption on test material can be calculated in the same manner, and the absorption rate per unit time per unit target mass is
\bea
R \sim \frac{1}{4}\frac{g_{aee}^2}{e^2 } {\omega^2 \over m_e^2}  \frac{\rho_a}{\rho_T} {\rm Im}\left[ - \frac{1}{\varepsilon(\omega)}\right]
\eea
where $\rho_a$ is the local energy density of the ALP. 

We show in the right-panel of the Fig.~\ref{figure2} the exclusion limit on the $g_{aee}$, where the shaded region is excluded. Here we do not consider the axion coming from the sun, and  take $\rho_a\sim 0.4 ~{\rm GeV/cm^3}$ as the local energy density of the ALP on the earth.  
For the mass range that we care, there are other constraints from solar neutrinos~\cite{Giudice:2012ms}, red giants~\cite{Gondolo:2008dd},  as well as solar axion measurement at the  XENONnT and the PandaX4T~\cite{XENON:2022ltv,PandaX-4T:2021bab}, which have $g_{aee}<2.82\times 10^{-11}, <1.9\times 10^{-12}$ and $<4.38\times 10^{-12}$, respectively. These constraints are much stronger than that from the current measurement result of the transmon qubit, so we only show the constraint of the electron g-2 in the plot, since it has encompassed all the parameter space detectable by the transmon qubit.
Although this exclusion limit is not stronger than limits from other observations, it provides the first direct detection limit on the ALP-matter coupling in the meV-eV mass range. Further experiments with more precise measurement and more systematic estimation of background might enable us to explore more competitive parameter space. 

Before proceeding to the discussion, Let's comment on formulas used in our analysis.  We acknowledge that the standard Lindhard/Mermin formalism used in DarkELF does not automatically incorporate the complex effects of superconductivity, such as the formation of a gap in the density of states. Our approach  provides a conservative, order-of-magnitude estimate for the absorption rate for dark matter. A more precise treatment is beyond the scope of the current version of DarkELF. This simplification is noted as an area for future refinement in more dedicated detector studies.  Furthermore, The derivation  of the absorption rate is based on some specific assumptions following the strategy of Ref. [40].  It works for current magnitude estimation of constraints,  and needs further dedicated calculation  when we intend to test a specific ultralight DM model using a transmon qubit device.

For the device performance, there are fundamental and readout related noise sources as noted in Ref. [73], such as telegraph noise, Fano noise, system and amplifier noise, etc.  We briefly address mechanisms for  mitigating noises. For the electromagnetic noise,  one may use high-conductivity materials for Faraday cages to suppress external radio frequency interference, employ coaxial shielding or cryogenic coaxial cables for signal lines to minimize crosstalk and radiation noise, use differential signal transmission to enhance robustness against common-mode electromagnetic interference.  For the thermal noise, one may employ dilution refrigerators to cool temperatures to the mK range, suppressing thermally excited quasiparticles,  design quasiparticle trap structures to accelerate recombination of thermally generated quasiparticles, use quantum impedance matching to compress thermal noise spectral density to the quantum limit.

\section{Discussion}

Ultralight bosonic particles have emerged as compelling candidates for dark matter (DM) due to their potential to address various cosmological and astrophysical mysteries. Detecting these elusive particles, however, presents a significant challenge to the scientific community.
%
%
Recent studies and technological advancements are pushing the boundaries of what is possible. The use of quantum technologies in various cavity experiments, antenna experiments and collective excitation experiments, etc., and the exploration of novel detection methods are paving the way for potentially groundbreaking discoveries in our understanding of DM.
Recent studies showed that quantum qubit, a foundation for exploring quantum computing,  can serve as competitive devices for direct detection of light DM. In this paper, we explored the constraints on light bosonic DM absorption from the measurement of quasiparticles in the transmon qubit. Our result gives the first constraints on dark photon, scalar and axion-like particle dark matter from existing transmon qubit measurements. This underscores the growing interest and progress in quantum technologies to play a role in the search for dark matter. 

In prospect, developments in quantum qubits can provide a promising avenue for light bosonic dark matter direct detection with more dedicated R\&D. In this paper the projective limit is a proof-of-principle estimate based on the sensitivity to the equilibrium density of quasi-particles in aluminum. Meanwhile, new detectors utilizing the fundamental architecture of transmon qubits are quickly advancing toward single Cooper-pair breaking sensitivity~\cite{Fink:2023xel,Li:2024xel,Ramanathan:2024xel}, demonstrating a notable reduction in the detection threshold and a substantial enhancement in sensitivity. When single quasi-particle sensitivity becomes a reality, direct measurement of the DM-induced production rate with a larger test mass and over longer exposure periods will likely improve the detection sensitivity significantly.

\begin{acknowledgments}
This work was supported in part by the National Key R\&D Program of China under Grant No. 2023YFA1607104, by the National Natural Science Foundation of China (11775025, 12150010, 12175027 and 12447105), and by the Fundamental Research Funds for the Central Universities under Grant No. 2017NT17, and by the State Key Laboratory of Particle Detection and Electronics Fund under Grant No. SKLPDE202310.
\end{acknowledgments}

\bibliography{references}

\end{document}